\documentclass[11pt]{article}
\usepackage{psfig,amsmath,amssymb}

\begin{document}

\renewcommand{\thefootnote}{\alph{footnote}}

\title{Mixing in the $D^0$ System \\  Results from Collider Experiments}

\author{Monika Grothe\thanks{E-mail: Monika.Grothe@cern.ch}
\and {\small \it European Organization for Nuclear Research CERN, EP division}
\and {\small \it 1211 Geneva 23, Switzerland}}

\date{}
\maketitle

\begin{abstract}
Mixing in the $D^0$ system may provide a sensitive probe for 
new physics beyond the Standard Model (SM) but has so far eluded
experimental observation. 
The SM predictions are typically small ($< 10^{-3}$) for the 
mixing parameters $x$, $y$ which, in the absence of charge-parity ($CP$) 
symmetry violation, measure
the mass ($x= \Delta m/ \Gamma$) and lifetime 
($y= \Delta \Gamma / 2 \Gamma$) difference of the $CP$ eigenstates in the
$D^0$ system. The asymmetric $B$-factory experiments BABAR and Belle open up 
the opportunity of measuring $x$, $y$ with unprecedented statistical precision
and sample purities. Results from BABAR and Belle, and 
from CLEO are reviewed.
\end{abstract}

\vspace*{-13cm}
\begin{flushright}
\small CERN-OPEN-2003-002
\end{flushright}

\newpage

\tableofcontents

\newpage

\section{Introduction}
\noindent
Mixing phenomena, i.e. the oscillation of a neutral 
meson into its corresponding anti-meson 
as a function of time,
have been observed in the $K^0$ and $B^0$ systems \cite{PDG}. Up to now, 
oscillations in the $D^0$ system have eluded experimental detection. 

This is in keeping with the Standard Model (SM) prediction that 
mixing in the $D^0$ system should be substantially smaller than for the 
$K^0$ and $B^0$ \cite{ligeti,petrov}. 
The SM prediction, as discussed below, is a consequence of 
$D^0$ - $\overline{D^0}$ mixing
being dominated by intermediate states with light $d$ and $s$ quarks, 
a feature unique to the $D^0$ system among the neutral mesons.

Mixing phenomena are characterized by two dimensionless parameters $x$, $y$. 
When charge-parity ($CP$) symmetry holds, they correspond to 
the difference in mass ($x= \Delta m/ \Gamma$) and lifetime
($y= \Delta \Gamma / 2 \Gamma$) of the $CP$ eigenstates in the neutral meson 
system. Calculations within the SM typically predict that $|x|, |y| < 10^{-3}$ 
for the $D^0$ \cite{nelson_comp}.
Due to the smallness of the SM prediction, 
mixing in the $D^0$ system may provide a sensitive probe for physics beyond 
the SM.

Independent of their discovery potential, precise measurements of $x$ and $y$
in the $D^0$ system are desirable because even SM predictions for $x$, $y$ span
several orders of magnitude \cite{nelson_comp}. The main challenge in SM
calculations of $D^0$ - $\overline{D^0}$ mixing is estimating the size of 
SU(3) flavor-symmetry breaking effects (see below) \cite{petrov}.

In this article, measurements of the mixing parameters $x$, $y$
in the $D^0$ system are reviewed. Results on $CP$ violation are only 
discussed when they arise as integral part of the $x$, $y$ measurements.

The most recent results come from the 
asymmetric $B$-factory experiments BABAR and Belle, in operation since
1999. BABAR, at the 
storage ring PEP-II of the Stanford Linear Accelerator Center (SLAC), USA,  
and Belle, at the KEK $B$-factory in Tsukuba, Japan, have accumulated the 
largest currently available charm samples 
(integrated luminosity $\sim 100 \ {\rm fb}^{-1}$ by the end of 2002) and 
can be expected to 
continue data taking till the end of the decade. Hence, BABAR and Belle open 
up the 
opportunity of studying charm decays with unprecedented statistical precision.

The first collider experiment to employ the methods for measuring $x$, $y$ 
now in use at BABAR and 
Belle was the CLEO experiment at the Cornell Electron Storage Ring (CESR), USA,
in operation until 1999.
This review describes these methods and the results obtained by CLEO, BABAR 
and Belle. Wherever appropriate, comparisons are made with 
results from fixed-target experiments, notably from E791 and FOCUS,  
that were dedicated to charm studies during the 1991/1992 (E791) and 
1996/1997 (FOCUS) fixed-target runs at Fermilab.

\section{Mixing formalism}
\noindent
Mesons are produced in strong and electromagnetic interactions as flavor 
eigenstates with a well-defined quark content. The production Hamiltonian 
is of the form 
${\mathcal H}_0 = {\mathcal H}_{strong} + {\mathcal H}_{em}$. Its 
eigenstates are the flavor or interaction eigenstates $M^0$, 
$\overline{M^0}$. The $CPT$ theorem requires that particle and anti-particle
have the same mass and lifetime. 
Thus $M^0$ and $\overline{M^0}$ 
correspond to identical eigenvalues $m_0$, $\gamma_0$ of ${\mathcal H}_0$.

Mesons are observed by way of their decays which are governed by the
weak force. The eigenstates of the evolution Hamiltonian that is responsible 
for their decay, 
${\mathcal H} = {\mathcal H}_0 + {\mathcal H}_{weak}$,
are the mass or decay eigenstates $M_{1,2}$ with the corresponding
eigenvalues $m_{1,2}$ and $\gamma_{1,2}$. 
They are the physically observable states that obey 
an exponential decay law with a slope given by the
respective lifetime $\gamma_{1,2}$:
\begin{equation}
 | M_{1,2} (t) \rangle = 
| M_{1,2}(0) \rangle e^{-im_{1,2}t} e^{-(\gamma_{1,2}/2) t}.
\end{equation}

As a consequence of the difference between production and evolution 
Hamiltonian, a sample of neutral mesons, produced, e.g., as a pure 
$M^0$ sample, evolves in time as a superposition of $M^0$ and 
$\overline{M^0}$ states with time-dependent coefficients: 
\begin{equation}
| M(t) \rangle = c_{M^0}(t) | M^0 \rangle + c_{\overline{M^0}} (t) 
| \overline{M^0} \rangle .
\end{equation}
The time evolution obeys Schr\"odinger's equation:
\begin{equation}
\frac{\partial}{\partial t} \left( \begin{array}{c} c_{M^0}(t) \\  
c_{\overline{M^0}}(t) \end{array} \right) =
-i {\mathcal H} \left( \begin{array}{c} c_{M^0}(t) \\  
c_{\overline{M^0}}(t) \end{array} \right) 
\ \ {\rm with} \ {\mathcal H} = 
\left[ \begin{array}{cc} m_0 - i \frac{\gamma_0}{2} & m_{12} - i \frac{\Gamma_{12}}{2} \\
m_{12}^\star - i \frac{\Gamma_{12}^\star}{2} & m_0-i \frac{\gamma_0}{2} \end{array} \right] .
\end{equation}
This equation represents two coupled differential equations.
Its eigenstates are the physically observable 
eigenstates $M_{1,2}$.
As a consequence of the non-zero off-diagonal elements in the Hamiltonian 
${\mathcal H}$,
the mass eigenstates $M_{1,2}$ are obtained as linear superpositions of the 
flavor eigenstates:
\begin{equation}
| M_{1,2} \rangle  = p | M^0 \rangle \pm q | \overline{M^0} \rangle.  
\end{equation}
\nopagebreak 
In the absence of $CP$ violation, i.e. for $|q/p| = 1$, $M_{1,2}$ are 
$CP$ eigenstates.

For a more detailed discussion, see Ref.~\cite{waldi}.

\subsection{The mixing parameters $x$, $y$}
\noindent
Mixing phenomena can be characterized by two dimensionless parameters:
\begin{equation}
 y = (\gamma_1 -\gamma_2) / (\gamma_1 + \gamma_2) = \Delta \Gamma / 2\Gamma,
 \quad \quad 
x = (m_1 - m_2) / \Gamma = \Delta m / \Gamma.
\end{equation}

Mixing in a neutral meson system occurs when at least one of the two 
following possibilities
applies: Between the decay eigenstates of the system there is a non-zero 
lifetime difference ($y \neq 0$), or there is a non-zero mass difference
($x \neq 0$). In the first case, flavor mixing is a consequence of the 
shorter-living
decay eigenstate dying out. The remaining longer-living one is a linear 
combination of a $M^0$ and a $\overline{M^0}$ component, so that even an 
initially pure $M^0$ sample shows some fraction of $\overline{M^0}$ 
after some time. In 
the second case, flavor mixing is a consequence
of a pure transition from $M^0$ to $\overline{M^0}$ and vice versa.    

In the $D^0$ system, according to the SM, $CP$ violation is a negligible 
effect (see below); 
the decay eigenstates are, to a good approximation, also $CP$ 
eigenstates. In the following, the eigenstate $D_1$ ($D_2$) with mass $m_1$ 
($m_2$) and width $\gamma_1$ ($\gamma_2$) is chosen as $CP$-even 
($CP$-odd).\footnote{This convention follows the one used by the CLEO
collaboration \cite{CLEO_xy}.}
In the $K^0$ and $B^0$ systems, where $CP$ violation is experimentally 
observable~\cite{cpviolKB}, the conventional choice for $K_1$ and $B_1$ 
($K_2$ and $B_2$) is the heavier (lighter) eigenstate. 

In oscillation plots that show the fraction of $\overline{M^0}$
flavor states in an initially pure sample of $M^0$ as a function of time, 
the parameter $x$ is 
related to the frequency of the oscillation, while $y$ is related to the 
damping of its amplitude.

\begin{figure}[htbp] 
\centerline{\psfig{file=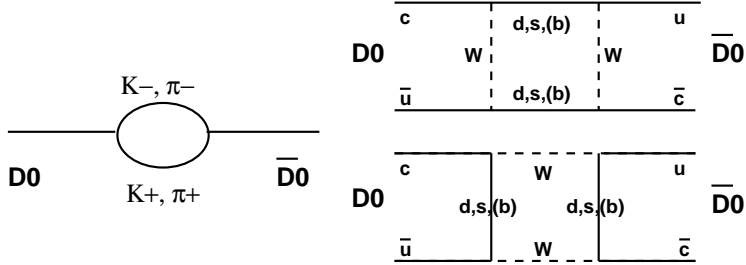,height=3.5cm}} 
\caption{\small The types of processes through which mixing can proceed in 
lowest order in the SM. 
Left: Long range processes with amplitude $\varpropto -iy$. Right: 
Short range processes with amplitude $\varpropto x$ \cite{CLEO_xy}.}
\label{fig:feynman}
\end{figure}

The two parameters $x$, $y$ reflect different mechanisms through which mixing 
can proceed in lowest order in the SM \cite{CLEO_xy}. 
A $M^0$ can oscillate into its anti-particle by way of 
on-shell intermediate states that are accessible to both particle and 
anti-particle. These are long-range processes with amplitudes $\propto -i y$. 
On the other hand, a $M^0$ can also oscillate 
by way of off-shell intermediate states that can be represented by
box-diagram loops (see Fig.~\ref{fig:feynman}). 
These are short-range processes with amplitudes $\propto x$.
Additional contributions to the box diagrams in Fig.~\ref{fig:feynman}(right) 
from as-of-yet 
unobserved non-SM particles could result in a deviation of the 
measured value of $x$ from the SM prediction.

The time-integrated probability of a neutral meson to oscillate first and then decay compared to decaying directly is described by the mixing rate $R_{mix}$.
In the absence of $CP$ violation in mixing, $R_{mix}$ is identical for $M^0$
and $\overline{M^0}$. The following expression then holds for the decay of a 
$M^0$ into
a final state $f$~\cite{waldi}:
\begin{equation}
R_{mix} = {\mathcal B} (M^0 \rightarrow \overline{M^0} \rightarrow \bar{f}) 
\ / \ {\mathcal B} (M^0 \rightarrow f) = 
(x^2 + y^2) / (2 + x^2 -y^2) .
\label{eq:Rmix}
\end{equation}

In the $D^0$ system, the SM expectations for $|x|$ and $|y|$ are typically 
below $10^{-3}$, which
results in $R_{mix}(D^0) < 10^{-6}$.
This explains why up to now no direct observation of flavor mixing in the 
$D^0$ system has been achieved and why such an observation is likely to be 
unaccessible also in the 
future if no non-SM effects come into play. 

The situation is much more favorable in the $K^0$ and $B^0$ systems.
Experimentally~\cite{PDG}, $x$ and $-y$ are known to be of the order unity for 
the $K^0$ system, which implies $R_{mix}(K^0) \sim 1$. 
In the $B^0$ system, $x \sim 0.7$
experimentally~\cite{PDG}, while $y/x \sim -1/500$ is expected~\cite{waldi}, 
resulting in
$R_{mix}(B^0) \sim 0.2$.\footnote{For the $B^0_s$ system, only upper limits 
have been determined experimentally \cite{PDG}: $x > 19.0$ at 95\% CL, while again
$y/x \sim -1/500$ is expected~\cite{waldi}, so that $R_{mix}(B^0_s) \sim 1$.
Direct experimental observation of mixing in the $B^0_s$ system is 
complicated by the very high oscillation frequency $\propto x$.}

\subsection{Predictions for mixing in the $D^0$ system}
\noindent
Predictions for $x$ and $y$ both within and beyond the SM span several orders 
of magnitude. For a compilation of predicted values, see 
Ref.~\cite{nelson_comp}. SM predictions give typically
$|x|,|y| < 10^{-3}$ and can be as low as $10^{-8}$. 
Contributions of 
physics beyond the SM may enhance $|x|$ to values of up to $10^{-2}$, while 
$|y|$ is assumed to be dominated by SM effects.

At the source of the large uncertainties in the theoretical description, 
even within the SM, is the charm mass being intermediate between heavy and 
light. The charm quark is too light for perturbative treatments to work well
but, on the other hand, it is too heavy for its decays to be dominated by a 
small number of final states. For a detailed discussion, see 
Ref.~\cite{ligeti,petrov}.

In the SM, the $D^0$ has the unique feature that its mixing proceeds via 
intermediate states of down-type quarks in the box diagrams of 
Fig.~\ref{fig:feynman}. 
The contribution of the bottom quark to the loops can be neglected 
because of the smallness of~\cite{golowich} $V_{ub} \times [(m_b^2-m_d^2)^2/m_W^2m_c^2]$. 
The first factor,
$V_{ub}$, is the coupling between quarks of the first and third generation
as described by the Cabibbo-Kobayashi-Maskawa (CKM) matrix. The second factor 
corresponds to GIM-type suppressions in the loop (see below). 
Consequently, the $D^0$ 
system can be treated to a good approximation as a mere two-generation problem
that the SM describes with the help of a $2 \times 2$ rotation matrix.
There, the rotation angle is the Cabibbo angle $\theta_C$, and no $CP$ violating
parameters are foreseen.

In that framework, the $D^0$ decay is 
Cabibbo-favored, while the box diagrams of Fig.~\ref{fig:feynman}(right) 
are doubly Cabibbo-suppressed. Thus in the $D^0$
system, the SM disfavors mixing as expressed by $x$ at least to the level 
$\sin{\theta_C}^2 \approx 0.05$. In the SU(3) flavor limit, where 
for the quark masses holds: $m_u=m_d=m_s$,
the sum over the box diagrams in Fig.~\ref{fig:feynman}(right) is zero 
(GIM-type cancellation\footnote{GIM cancellation led Glashow, 
Illiopoulos and Maiani to infer from the measured small size of the ratio
$BR(K_L^0 \rightarrow \mu^+ \mu^-)/BR(K_L^0 \rightarrow {\rm all \ modes})$ 
the existence 
of a fourth quark, the charm quark.}) and $x$ would be zero accordingly.
In the SM, where $SU(3)$ flavor symmetry is broken, GIM-type cancellations
result in suppression factors that depend on the mass of the heaviest quark
in the loop relative to the $W$ boson mass: 
$(m_s^2-m_d^2)^2/m_W^2m_c^2 \sim m_s^2/m_W^2$ for the $D^0$,
as compared to $\sim m_t^2/m_W^2$ ($\sim m_c^2/m_W^2$) for the $B^0$ ($K^0$).
    
For the on-shell intermediate states of Fig.~\ref{fig:feynman}(left), which
correspond to mixing in the $D^0$ system as expressed by $y$,
light quark states are favored by phase
space. While the limited available phase space for decays results in a
large lifetime difference in the $K^0$ system ($y \sim 1$), such 
constraints for decays in the $D^0$ or $B^0$ system are much weaker. 
The measured branching ratios for $D^0$ decays, e.g. into $\pi^+ \pi^-$, 
$K^+K^-$ or 
$K_s^0 \pi^0$ are small and not comparable to the ones in the $K^0$ system. 
Based on the same GIM-type suppression arguments as for $x$, SM predictions
typically arrive at $|y| < 10^{-3}$. However, as for $|x|$, the precise size of
these GIM-type suppressions is a major source of uncertainty. Recent 
theoretical work~\cite{ligeti,petrov} points out the possibility that 
SU(3) flavor breaking may 
result in $|y| \sim 0.01$ being natural in the SM. Such a large value of
$|y|$ would lower the sensitivity of experimental measurements of
$D^0$ mixing to potential non-SM physics effects in $|x|$.

\section{Methods for determining $x$, $y$ in the $D^0$ system}
\noindent
Given the smallness of the expected mixing rate, $R_{mix}$, in the $D^0$ 
system, the 
extraction of the mixing parameters from a direct observation of flavor 
oscillations as a function of time is not feasible. Instead,
three complementary methods have been
applied by experiments:\footnote{The charge conjugate 
modes are always implied unless stated otherwise. The experiments treat 
$D^0$ and $\overline{D^0}$ separately only when investigating the possibility
of $CP$ violating effects.}
\begin{itemize}
\item[1.]
By comparing the lifetimes measured in hadronic $D^0$ decays of
specific $CP$ symmetry, $y$ can be determined. 
\item[2.]
Measurements using the wrong-sign $D^0$ decays: 
\begin{itemize}
\vspace*{-0.15cm}
\item[(a)]
From the time evolution of wrong-sign hadronic decays, e.g.
$D^0 \rightarrow K^+ \pi^-$, $x^{\prime 2}$ and $y^{\prime}$ can be determined
independently. $x^\prime$, $y^\prime$ are related to $x$, $y$ by a rotation. 
\item[(b)]
From wrong-sign semileptonic decays, e.g. 
$D^0 \rightarrow K^+ \l^- \bar{\nu_l}$, the linear
combination $x^2 + y^2$ can be determined.  
\end{itemize}
\end{itemize}

In order to compare directly the results of method 2(a) to those of methods
1, 2(b), an assumption is necessary for the size of the rotation angle 
(see below) between $x$, $y$ and $x^\prime$, $y^\prime$. 

The results of BABAR, Belle and CLEO were obtained with the first
two methods. They are discussed in detail in the following. 
Comparisons to earlier results by FOCUS (methods 1, 2(a)) and E791 
(all three methods) are given where appropriate. 
 
BABAR, Belle and CLEO use for their measurements the following hadronic 
two-prong decays of the $D^0$:
\begin{itemize}
\item The Cabibbo-favored (CF), so-called right-sign (RS) decay 
$D^0 \rightarrow K^- \pi^+$.
\item The $CP$-even decays $D^0 \rightarrow K^- K^+$ and 
$D^0 \rightarrow \pi^- \pi^+$ which are singly Cabibbo-suppressed. The former
(latter) occurs $\sim 1/9$ ($\sim 1/25$) times less frequently than the RS 
decay.\footnote{In the limit of SU(3) flavor
symmetry, the branching ratios for the $D^0$ decays into $K^-K^+$ and 
$\pi^- \pi^+$ are identical. The experimentally observed substantial difference
is an example of large SU(3) flavor breaking effects in the $D^0$ 
system \cite{nir}.}
\item The so-called wrong-sign (WS) decay  
$D^0 \rightarrow K^+ \pi^-$ which, in the absence of mixing, is 
doubly Cabibbo-suppressed (DCS) and occurs  $\sim 1/300$ times 
less frequently than the RS decay.
\end{itemize}

\subsection{Measurement of $y$ from a $D^0$ lifetime ratio}
\noindent
In the absence of $CP$ violation, the final state of the RS $D^0$ decay
is an equal mixture of $CP$-even and $CP$-odd states. 
Under the assumption that $x$ and 
$y$ are small, the
decay-time distribution is approximately exponential~\cite{nir} with a slope
$\tau_{RS} = 2/(\gamma_1 + \gamma_2)$, where $\gamma_1$ ($\gamma_2$) is the 
width of the $CP$-even ($CP$-odd) decay eigenstate $D_1$ ($D_2$) defined in 
Sec.~2. 
The decay-time distributions for the $D^0$ decays into the
$CP$-even final states $K^-K^+$ and $\pi^- \pi^+$ are exponential with slope
$\tau_{KK}= \tau_{\pi \pi} = 1/ \gamma_1$. 

Then, in the absence of $CP$ violation, 
$y=\Delta \Gamma /2 \Gamma = y_{CP}$ with:
\begin{equation}
\quad \quad  y_{CP} = \frac{\tau_{RS}}{\tau_{KK}} -1  \quad \quad {\rm or} \quad \quad 
y_{CP} =\frac{\tau_{RS}}{\tau_{\pi \pi}} -1 .  
\label{eq:ycp}
\end{equation}
In the available measurements $y = y_{CP}$ is assumed.\footnote{ 
If $CP$ violation is present, then $y \neq y_{CP}$ and
$y_{CP} = y \cos \phi - 0.5 x A_M \sin \phi$, 
where $A_M, \phi$ are $CP$ violating parameters related to $CP$ violation in 
mixing ($A_M$) and in the interference of decays with (CF) and without (DCS) 
mixing
($\phi$) \cite{nir}.}

\subsection{Measurement of the time evolution of hadronic WS $D^0$ decays} 
\label{sec:WS}
\noindent
The $D^0$ can arrive at a WS hadronic final state in two ways, either by
undergoing directly the DCS decay or by first 
oscillating into a $\overline{D^0}$ that then undergoes a CF 
decay. This gives rise to three different components in the 
WS decay: from the DCS decay, from mixing and from the interference of the 
decays with (CF) and without (DCS) mixing.

Assuming $CP$ conservation and expanding the decay rate 
up to ${\mathcal O} (x^2)$, ${\mathcal O} (y^2)$ results in the following 
approximation~\cite{CLEO_xy,nir} for the time evolution\footnote{This formula is valid for $t \lesssim \Gamma$.}  of the 
hadronic\footnote{Semileptonic WS $D^0$ decays can only arise
via mixing, so that $\Gamma_{WS}^{sl}(t) \ \propto \  
\exp(-t)( \ x^{2} \ + y^{2} \ ) \ t^2$.} 
WS decay rate:
\begin{equation}
\Gamma_{WS}(t) \ \propto \  \exp(-t) \ [ \ R_D \ + \ \sqrt{R_D} \ y^\prime \ t \ + \ 1/4 \
( \ x^{\prime 2} \ + y^{\prime 2} \ ) \ t^2 \ ],
\label{eq:mix}
\end{equation}
while for the RS decays: $\Gamma_{RS}(t) \ \propto \exp(-t)$.
Here, $t$ is given in units of the lifetime of the $D^0$.
In this approximation, the time-independent coefficient $R_D$ corresponds to the 
DCS component.
Even if the mixing contribution, quadratic in $t$, 
is very small, the interference term, linear in $t$, may result in a 
discernible deviation from the exponential time evolution characterizing a 
pure decay.

The parameters $x^\prime$
and $y^\prime$ are related to the mixing parameters $x$, $y$ by a rotation:
\begin{equation}
x^\prime \ = \ x  \ \cos{\delta_{K \pi}} \ + y \ \sin{\delta_{K \pi}}, \quad \quad
y^\prime \ = \ y  \ \cos{\delta_{K \pi}} \ - x \ \sin{\delta_{K \pi}}.
\label{eq:phase}
\end{equation}
The phase, $\delta_{K \pi}$, is a strong phase between the DCS
contribution and the CF one and does not 
violate $CP$ symmetry.

In a measurement based on the time-evolution of the WS rate alone, it is not
possible to determine the phase, $\delta_{K \pi}$.
Furthermore, since only the
square of $x^\prime$ enters into $\Gamma_{WS}(t)$, the sign of $x^\prime$ 
cannot be determined, either.

If there is $CP$ violation in the $D^0$ system, then the parameters $R_D$, 
$x^\prime$, $y^\prime$ in Eq.~\ref{eq:mix} have to be substituted by:
$R_D^\pm = \sqrt{(1 \pm A_D)/(1\mp A_D)} R_D$,
$x^{\prime \pm} = \sqrt[4]{K^\pm} (x^\prime \cos{\phi} \pm y^\prime \sin{\phi})$, 
$y^{\prime \pm} = \sqrt[4]{K^\pm} (y^\prime \cos{\phi} \mp x^\prime \sin{\phi})$, where $K^\pm =(1 \pm A_M)/(1\mp A_M)$.
The plus (minus) sign pertains to the decay of a $D^0$ 
($\overline{D^0}$).
The three additional parameters are related to $CP$ violation in the
DCS decay ($A_{D}$), in the mixing term ($A_M$) and 
in their interference (phase $\phi$) \cite{nir}. 

CLEO~\cite{CLEO_xy} in its analysis uses a somewhat different approximation 
which is valid for $A_D, A_M \ll 1$:
$R_D^\pm = (1 \pm A_D) R_D$,
$x^{\prime \pm} = \sqrt{1 \pm A_M}(x^\prime \cos  \phi \pm y^\prime \sin \phi)$, 
$y^{\prime \pm} = \sqrt{1 \pm A_M} (y^\prime \cos  \phi \mp x^\prime \sin \phi)$.

The total time-integrated hadronic WS rate, assuming $CP$ conservation, 
is:\footnote{In this approximation, the mixing rate (see Eq.~\ref{eq:Rmix})
appears as $R_{mix} =1/2 (x^{\prime 2} + y^{\prime 2}) = 1/2 (x^{2} + y^{2})$.}
\begin{equation}
R_{WS} \ = \ \int \Gamma_{WS}(t) \ / \ \int \Gamma_{RS} (t) \ = \
R_D \ + \ \sqrt{R_D} \ y^{\prime} \ + \ 1/2 \ (x^{\prime 2} + y^{\prime 2}). 
\end{equation}
If there is no mixing in the $D^0$ system, then $R_{WS}$ reflects 
only the rate of the DCS decay: $R_{WS} = R_D$. 
The SM predicts in the limit of SU(3) flavor symmetry:
$R_D \sim \tan^4 \theta_C \approx 0.0025.$
SU(3) symmetry breaking effects may increase this value \cite{nir}.

\subsection{Measurement of the strong phase $\delta_{K \pi}$}
\noindent
No experimental determination of the strong phase, $\delta_{K \pi}$, is yet
available.

Reference~\cite{phase} outlines a method to extract $\delta_{K \pi}$ from
a measurement of the rates of the DCS and CF decays of the type
$D \rightarrow K \pi$. This method requires the determination of rate 
asymmetries in $D$ meson decays to $K_L \pi$ and $K_S \pi$. In 
Ref.~\cite{belle_delta} Belle finds that the relevant measurements of $K_L$ and
$K_S$ mesons are possible with a statistical precision sufficient to
constrain $\delta_{K \pi}$ \cite{yabsley}.

\section{The experiments}
\noindent
The experiments BABAR and Belle at the asymmetric B-factories PEP-II and KEK-B 
and CLEO at the storage ring CESR operate near the $\Upsilon(4S)$ resonance
at a center-of-mass energy of 10.6~GeV.
PEP-II and KEK-B were designed
with the primary goal of serving as asymmetric B-factories for the study
of $CP$ violation in the $B^0$ system. 
In the production cross section, $\sigma(e^+e^- \rightarrow q \bar{q})
\approx 4.45$~nb at $\sqrt{s} \approx 10.6$~GeV, $b \bar{b}$ 
final states account for~\cite{BABARphysbook} 
$\sim$~1.05~nb and $c \bar{c}$ final states for 
$\sim$~1.30~nb. In terms of production cross sections, the 
B-factories therefore serve equally well as charm-factories. 

Between October 1999 and October 2002,
BABAR recorded 94~${\rm fb}^{-1}$ of data, while Belle recorded 
98~${\rm fb}^{-1}$. 
The data set analyzed by CLEO,
recorded between February 1996 and February 1999, corresponds to 
an integrated luminosity of 9.0~${\rm fb}^{-1}$. 
The largest
available charm data sets obtained in fixed-target experiments were collected
at Fermilab by the heavy-flavor photoproduction experiment FOCUS and the
heavy-flavor hadroproduction experiment E791. These data sets  
contain $\sim 120,000$ (FOCUS)~\cite{focus_y} and $\sim 35,000$ 
(E791)~\cite{e791_y} 
identified $D^0 \rightarrow K^- \pi^+$ candidate events, compared
to, for example, $\sim 260,000$ events~\cite{BABAR_y} in the BABAR data set. 

Detailed descriptions of the BABAR and Belle detectors can be found 
elsewhere \cite{BABARdet,belledet}.
CLEO carried out its measurement with the CLEO II.V detector \cite{CLEOdet}.
Of primary importance for the reconstruction and identification
of $D^0$ decays are the vertex and particle-identification detectors. 
All three experiments are equipped with 
double-sided silicon vertex detectors.
At BABAR, the Silicon Vertex Tracker (SVT) 
(five layers, innermost radius $r_{\min} = 3.2$~cm) is also capable of 
stand-alone track reconstruction down to particle momenta of $\sim 60$~MeV. 
Mounted just outside of the BABAR drift chamber (DCH) volume 
is the ring-imaging detector of internally reflected Cherenkov light 
(DIRC). 
At Belle, the Silicon Vertex Detector (SVD) 
(three layers, $r_{\min} = 3.0$~cm) and the central drift chamber 
are surrounded by an array of aerogel Cherenkov
and time-of-flight scintillation counters (TOF).  
The CLEO II.V Silicon Vertex Detector (SVX) had three layers, with 
$r_{\min} = 2.35$~cm.

Differentiating kaons from pions is crucial for distinguishing with high
purity between the different two-prong $D^0$ decays listed in 
Sec.~3.
CLEO relies primarily on kinematic selection cuts based on the measured 
momenta and the assigned mass to distinguish between the different $D^0$ decay 
modes \cite{CLEO_xy,CLEO_y}. BABAR and Belle 
enhance their particle-identification capabilities considerably with the help
of likelihoods determined from a combination of the energy-loss (dE/dx) 
in the tracking devices and Cherenkov detector
information. Based on the ratio of the likelihoods for the two particle 
hypotheses, BABAR reaches an average $K^\pm$ identification efficiency
(mis-identification probability) of
$> 75\%$ ($< 8\%$) for momenta up to 4~GeV \cite{BABAR_xy}. 
Belle includes also the TOF information in 
the likelihood calculation and reports an efficiency of $\sim 85$\% and a 
mis-identification probability of $\sim 10$\% for momenta up to 
3.5~GeV \cite{belle_y}.

\section{Measurement of $y$}
\noindent
Following Eq.~\ref{eq:ycp}, the mixing parameter $y$ is determined by measuring the
slope of the decay-time distributions in independent candidate samples of
$D^0 \rightarrow K^- \pi^+$ and the singly-Cabibbo suppressed channels 
$D^0 \rightarrow K^- K^+$ and $D^0 \rightarrow \pi^- \pi^+$. 

The selection of a $D^0$ candidate, the determination of its proper 
decay-time and the general structure of the fit to obtain the $D^0$ lifetime
from the proper decay-time distribution are discussed in some detail in the
following. They are of importance also for the measurements with WS events
described in Sec.~6.

\subsection{Method of measurement}
\label{sec:method}
\noindent
$D^0$ candidates are selected by searching for pairs of tracks with opposite 
charge and combined invariant mass near the expected $D^0$ mass.
The common vertex of the track pair determines the $D^0$ candidate decay 
vertex, $\vec{v}_{dec}$, with typical resolutions as listed in 
Tab.~\ref{tab:exp_param}.
The interception point of the $D^0$ momentum vector, $\vec{p}_D$,
with the envelope of the interaction point (IP) 
provides the production vertex, $\vec{v}_{prod}$, of the $D^0$ 
candidate \cite{BABAR_y,CLEO_y,belle_y}.  

\begin{table}[!t]
\caption{\small Comparison of parameters that enter into the calculation of the 
proper decay-time, $t$, at BABAR, Belle and 
CLEO \cite{CLEO_y,BABAR_xy,belle_y}.}
\centerline{\footnotesize }
\centerline{
\begin{tabular}{|c| c c c |}
\hline
& BABAR & Belle & CLEO \\
\hline
IP envelope & & & \\
vertical [$\mu$m] & 6  & 2--4 & 10 \\  
horizontal [$\mu$m] & 120  & 80--120 & 300 \\  
along beam-axis [mm] & 8 & 3--4 & 10 \\
\hline
resolution & 80 $\mu$m & 110 $\mu$m rms & 40 $\mu$m \\
$D^0$ decay vtx & along $\vec{p}_D$ & along $\vec{p}_D$ & in each dim. \\   
\hline
resolution of proper decay-time $\sigma_t$ [fs]    & 180 &  & $0.4 \times \tau_D$ \\
\hline
\end{tabular}}
\label{tab:exp_param}
\end{table}

The proper decay-time of a $D^0$ candidate is derived from its mass 
($m_D(PDG) = 1.864$~GeV)~\cite{PDG} and its flight length. In order to
take resolution effects properly into account, that can, e.g.,
result in reconstructing a negative value of $t$, the flight length is
calculated from the projection of $\vec{v}_{dec} - \vec{v}_{prod}$ onto 
$\vec{p}_D$:
\begin{equation}
c \ t = m_D \ [ \ (\vec{v}_{dec} - \vec{v}_{prod}) \ \cdot \ \vec{p}_D / | \vec{p}_D |^2 \ ].  
\label{eq:t}
\end{equation}
The size of the IP envelope enters into the error on the proper decay-time, 
$\sigma_t$, through the uncertainty on the production vertex. 
The IP envelope is smallest in the plane 
transverse to the beam, as shown in Tab.~\ref{tab:exp_param}.
The flight length of a $D^0$ candidate
is typically $\sim 200~\mu$m in BABAR and Belle. 
The higher $D^0$ decay vertex resolution achieved by CLEO reflects the absence 
of boost-related uncertainties since at CESR, the $\Upsilon (4S)$ is
produced at rest, different from PEP-II and KEK-B, where $\beta \gamma = 0.56$
and $\beta \gamma = 0.425$, respectively, along the beam direction.

\subsection{Event selection}
\noindent
The three experiments use a similar set of quantities to reduce 
backgrounds \cite{BABAR_y,CLEO_y,belle_y}. 
The main differences lie in whether a $D^\star$ tag is required and which  
requirements are imposed for particle identification.

BABAR and CLEO select events likely to contain a $D^0$ with the help of the
decay $D^{\star \pm} \rightarrow D^0 \pi^\pm$, where the $\pi^\pm$ has very 
low momentum (``slow pion'', $\pi_s$).
Each $\pi_s$ candidate track is refitted with the constraint to coincide 
with the $D^0$ candidate production vertex, determined as described
in Sec.~5.1. This reduces substantially 
the mismeasurement of the $\pi_s$ momentum caused by multiple 
scattering.
Then the difference in the 
invariant mass of the $D^\star$ and the $D^0$ candidate, $\delta m$, 
is required not to deviate from the nominal one by more than a certain margin.
In BABAR, this margin is 2 or 3~MeV, depending on whether the $\pi_s$ 
reached the DCH; in CLEO the margin is 1~MeV.  
Belle does not use a $D^\star$ tag and requires solely that the $D^0$ 
candidate flight path be consistent with originating at the IP.

All three experiments reject events with secondary charm production 
from $B$ meson decays by requiring a minimum center-of-mass momentum of the 
$D^\star$ or $D^0$ candidates 
(2.5~GeV for BABAR and Belle, 2.3~GeV for CLEO).

All three experiments reduce background from random combinations of two 
tracks forming a $D^0$ candidate with the help of the angle 
$\theta^\star$,
measured in the $D^0$ candidate center-of-mass system, between the direction 
of the $D^0$ ($\overline{D^0}$) candidate boost and its positive (negative) 
daughter. BABAR requires $\cos \theta^\star > -0.75$ for pion daughters; 
Belle requires $\cos \theta^\star > -0.85$ for pion and, in the 
$K^-K^+$ channel, $|\cos \theta^\star| < 0.9$ for kaon daughters; 
CLEO requires $|\cos \theta^\star| < 0.8$ in all cases.   
BABAR and Belle apply in addition their particle-identification algorithms 
to the daughters of the $D^0$ candidate. CLEO requires in addition 
that the invariant mass obtained with the other possible particle-type 
assignments for the two daughters should be more than four standard 
deviations away from the nominal $D^0$ mass.

\begin{table}[t!]
\caption{ \small Data samples used by BABAR, Belle and 
CLEO in their measurements of $y$ \cite{BABAR_y,CLEO_y,belle_y}.}
\centerline{\footnotesize }
\centerline{
\begin{tabular}{|c| c c c |}
\hline
& BABAR & Belle & CLEO \\
\hline
$\mathcal L$ [${\rm fb}^{-1}$] & 57.8 & 23.4 & 9.0 \\
\hline
$K^-\pi^+$ candidate events   & 158,000 & 214,000 & 20,300 \\
Purity [\%]                   & 99 & 87 & 91 \\
\hline
$K^-K^+$ candidate events     &  16,500 &  18,300 &  2,500 \\
Purity [\%]                   & 97 & 67 & 49 \\
\hline
$\pi^-\pi^+$ candidate events &   8,400 &    --    &   930 \\
Purity [\%]                   & 92 & -- & 71 \\
\hline
\end{tabular}}
\label{tab:sel}
\end{table}

The analyses by BABAR and Belle reported here refer only to a sub-sample 
of their full data sets. The sub-sample sizes and the number of events
after the full event selection are listed in Tab.~\ref{tab:sel}.
The purities quoted there refer to a window around the mean $D^0$ mass.
The window size is $\pm 20$~MeV in BABAR, $\pm 3 \sigma$ in Belle
($\sigma(K^-\pi^+) = 6.5$~MeV,
$\sigma(K^-K^+) = 5.4$~MeV)
and $\pm 40$~MeV in CLEO. 
Although BABAR's data sample corresponds to twice the integrated luminosity 
of Belle's, Belle's $D^0$ candidate event samples are larger than BABAR's
and have lower purities because Belle does not require
a $D^\star$ tag in its event selection. CLEO arrives at lower efficiencies 
and, for the $CP$-even channels, lower purities because CLEO, lacking
dedicated particle-identification detectors, relies primarily 
on kinematic selection cuts to distinguish between the different $D^0$ decay 
modes.

\subsection{Lifetime fit}
\label{sec:yfit}
\noindent
The three experiments determine $y$ from unbinned maximum-likelihood fits
to the distribution of the reconstructed proper decay-time, $t$, 
(see Eq.~\ref{eq:t}) of the 
$D^0$ candidates \cite{BABAR_y,CLEO_y,belle_y}. 
No background subtraction is performed before
carrying out the fits; rather, it is left to the fit to describe both
signal and background.

The three experiments proceed in a comparable way to define the full fit 
function:
For each of the $D^0$ decay channels, they define a likelihood
function as the sum of a signal and a background part.
Both parts consist of a decay-time distribution convolved with a 
resolution
function. The signal decay-time distribution is exponential with a slope 
that corresponds to an inverse lifetime. The background decay-time 
distribution consists of two components, an exponential and a delta function,
that model background events with non-zero and with zero 
lifetime, respectively.
The former can, for example, arise from only partially reconstructed 
three-prong $D^0$ decays; the latter corresponds to combinations
of particles 
into a $D^0$ candidate 
that in fact originate at the IP. 

The resolution functions need to accommodate correctly reconstructed 
events and also events with misreconstructed parameters. This need is 
typically met by adding one or more Gaussians to accommodate
events that are misreconstructed to varying degree. The width of the 
Gaussian(s) modeling the resolution of well reconstructed events is
typically given by the per-event error on the reconstructed proper time $t$, 
$\sigma_t$, multiplied with a proportionality factor. The use of $\sigma_t$ 
allows to take into account the uncertainty from the highly elliptical shape 
of the IP envelope (see Sec.~5.1) in the fit.

\begin{figure}[!t] 
\begin{tabular}{cc}
\begin{minipage}[b]{.48\textwidth}
\centerline{\psfig{file=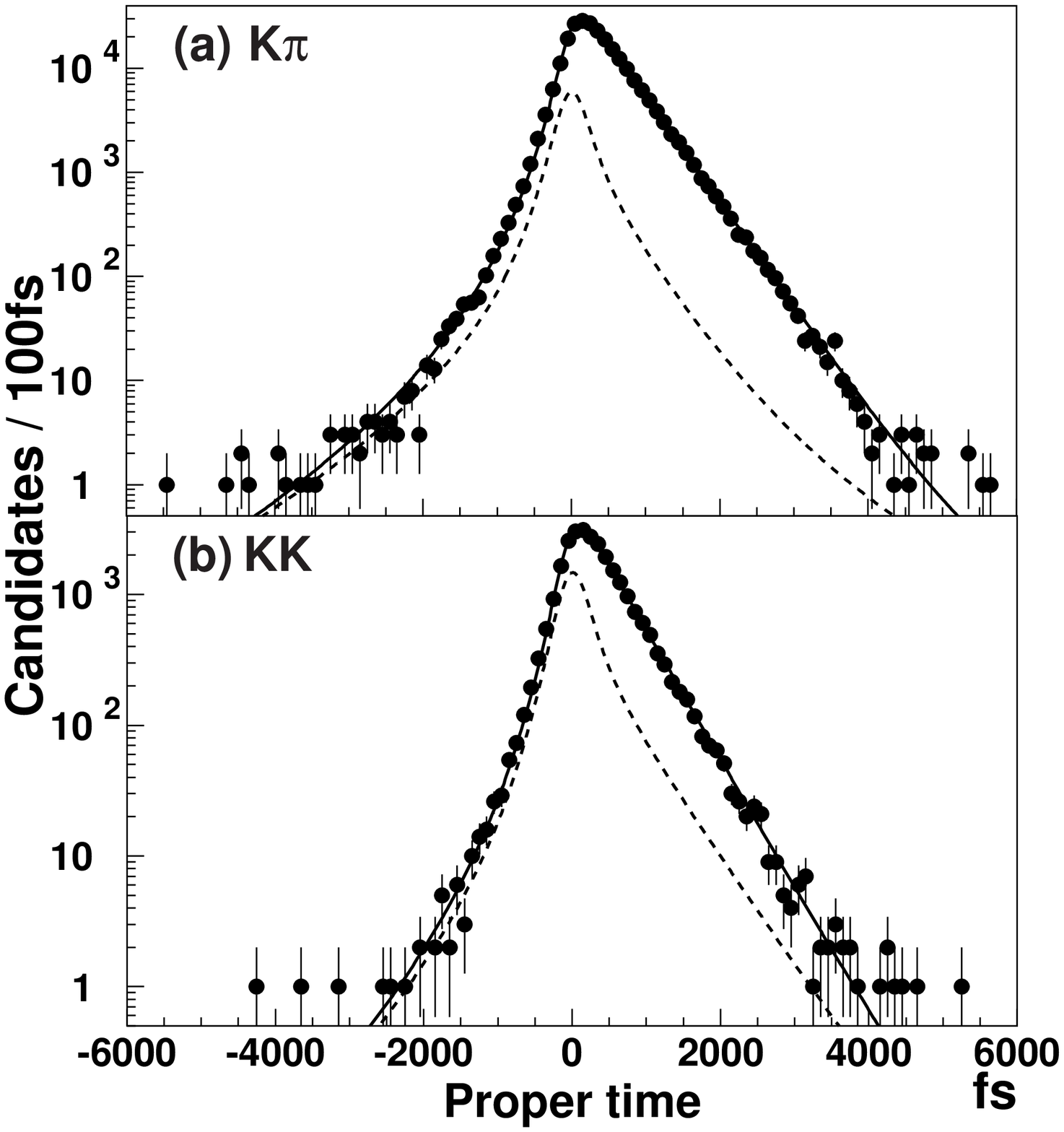,height=7.cm}} 
\end{minipage}
&
\begin{minipage}[b]{.48\textwidth}
\centerline{\psfig{file=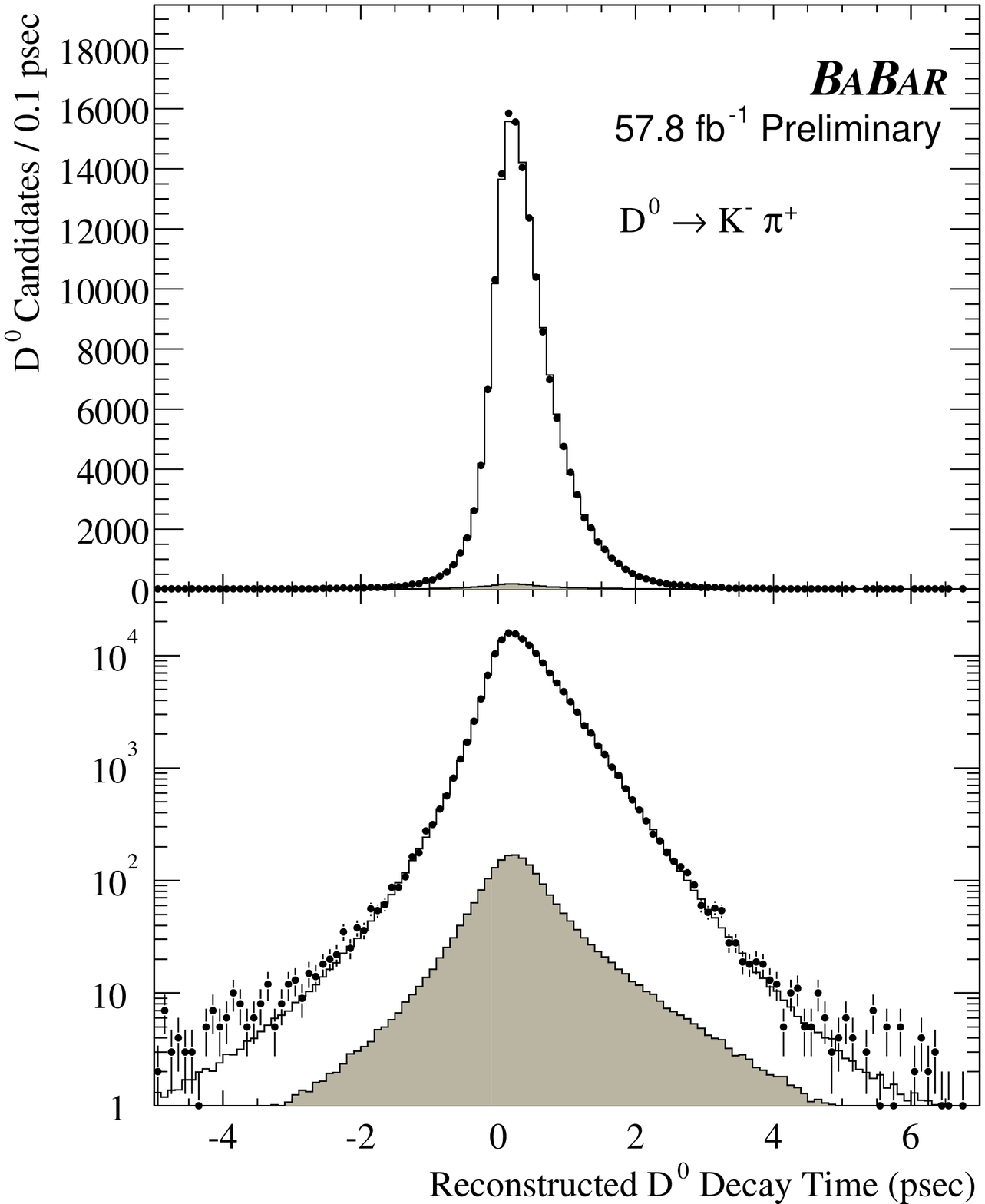,height=7.cm}} 
\end{minipage}
\end{tabular}
\caption{\small Proper decay-time distribution and fit result. The dots represent the
data points with statistical error bars. Left: Belle~\cite{belle_y}: 
For $D^0 \rightarrow K^- \pi^+$ (a) and $D^0 \rightarrow K^- K^+$ (b), in the 
$D^0$ mass signal region $\pm 3 \sigma$ around the mean $D^0$ mass. The solid
line is the result of the fit, the dashed line indicates the background
contribution. Right: BABAR~\cite{BABAR_y}: For $D^0 \rightarrow K^- \pi^+$, on 
a linear (upper) and logarithmic (lower) scale for all events including the 
$D^0$ mass sidebands. The solid line is the result of the fit, the gray area 
indicates the contribution attributed to background by the fit.}
\label{fig:lifetimefit}
\end{figure}
  
All three experiments determine the parameters of the background part
of the likelihood fit function
independently of those in the signal part. Also, all three experiments
enhance the ability of their fits to differentiate signal from background 
events by multiplying the likelihood function with a per-event probability to
be a signal event. The further away its reconstructed mass, 
$m_D$, is from the $D^0$ mass peak, the less likely is the $D^0$ candidate 
to belong to the signal. The probability is derived from an independent fit 
to the $m_D$ distribution in each $D^0$ decay channel.

The window in $m_D$ used in the fit must be chosen such that, in addition to
the $D^0$ signal peak, sufficiently wide sidebands are included so that the 
fit can extract information on shape and size of the background. 
BABAR uses mass windows of 140~MeV width, 
defined such that reflection peaks, e.g., from mis-identified $K^- \pi^+$ 
events in the $K^- K^+$ and $\pi^- \pi^+$ candidate samples, are excluded. 
Belle uses
a mass window of $\pm 40$~MeV around the mean and CLEO uses the window 
[1.825, 1.905~GeV]. The fit results for BABAR and Belle are shown in 
Fig.~\ref{fig:lifetimefit}. The tail at negative values in the $t$ 
distribution 
reflects the effect of the resolution. Comparing data and fit result 
in the unphysical region $t<0$
provides information about how well the resolution in the data is modeled
by the fit function. The distribution for $t>0$ reflects the resolution 
function convolved with the 
decay-time function.

\subsection{Discussion of systematic uncertainties and results}
\noindent
Belle~\cite{belle_y} determines $y$ from the lifetimes measured  
in $D^0$ decays into $K^-K^+$ and $K^- \pi^+$ ($y_{KK}$).
BABAR~\cite{BABAR_y} and CLEO~\cite{CLEO_y} use in addition the 
$\pi^- \pi^+$ channel ($y_{\pi\pi}$). 

\begin{figure}[tbp] 
\centerline{\psfig{file=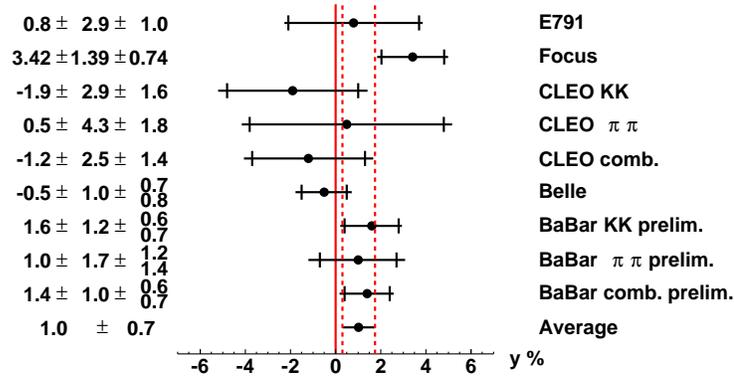,height=5cm}} 
\caption{\small Comparison of results for $y$ from different 
experiments \cite{focus_y,e791_y,BABAR_y,CLEO_y,belle_y}. The total size
of the error bars corresponds to the quadratic sum of statistical and 
systematic error, the inner part indicates the size of the statistical 
error only. The average is calculated as mean of the individual
measurements, each weighted by the quadratic sum of its statistical and
systematic error.}
\label{fig:y_comp}
\end{figure}

Belle and CLEO correct the lifetimes measured in the 
individual decay channels for small biases found with the help of Monte Carlo
(MC)
studies.
Belle reports that the proper decay-time tends to be reconstructed slightly
too small by a decay-channel dependent value.
The resulting correction to $y$ is $(-0.3 \pm 0.3)$\%. The 
error on this correction, due to the limited MC statistics, enters 
in the final systematic error on the measurement.
CLEO finds bias values in the individual lifetimes that are compatible,
within the statistical uncertainty on the MC bias estimate, with 
zero, chooses to apply corrections
nonetheless and includes their statistical uncertainty 
in the final systematic error ($\pm 1.0$\%
for $y_{KK}$, $\pm 1.4$\% for $y_{\pi\pi}$). BABAR also finds bias values
compatible with zero and
does not apply any corrections to the lifetimes but includes in the
systematic error on $y$ the statistical 
uncertainty on the MC bias estimate ($^{+0.4}_{-0.6}$\% for $y_{KK}$,
$^{+0.4}_{-0.9}$\% for $y_{\pi\pi}$).

Since $y$ is measured from the ratio of lifetimes, many systematic effects that
affect the individual lifetimes do not affect $y$. Both Belle and CLEO
find that the dominant remaining systematic error source is their 
understanding of the background contribution to the signals. 
In BABAR, the dominating source of systematic error is the above 
quoted uncertainty on the MC bias estimate.

All three experiments arrive at their final systematic error estimate
by summing in quadrature the contributions of the individual sources. 
BABAR and CLEO each determine $y$ as weighted mean of their measured
$y_{KK}$ and $y_{\pi\pi}$ values.

The results are listed in Fig.~\ref{fig:y_comp}.
Also shown are results from the fixed-target experiments\footnote{
The fixed-target experiment results are obtained by means of a fit
to the reduced proper time. This quantity includes a cut on a minimum 
distance between primary and secondary vertex as a principal 
tool to reduce non-charm background and is measured with a 
resolution of less than 1/10 of the $D^0$ lifetime. The FOCUS (E791)
result on $y$ is derived from $\sim 120,000$ ($\sim 35,400$) 
events in the $K^- \pi^+$ channel and $\sim 10,300$ ($\sim 3,200$)
events in the $K^-K^+$ channel.} 
 FOCUS~\cite{focus_y} and E791~\cite{e791_y}.
Within their errors, the results are compatible with each other and with the
SM expectation of a value of $|y|$ close to zero.

\section{Measurements with the WS decays $D^0 \rightarrow K^+ \pi^-$, 
$\overline{D^0} \rightarrow K^- \pi^+$}
\noindent 
As discussed in Sec.~3.2, the time evolution of the WS decay rate
provides a means of measuring simultaneously $x^{\prime 2}$ and $y^\prime$.
CLEO has published a result~\cite{CLEO_xy} obtained with 
$9.0 \ {\rm fb}^{-1}$ of data.
BABAR has presented a preliminary analysis~\cite{BABAR_xy} based on a data 
sample of 
$57.1 \ {\rm fb}^{-1}$.
At Belle, work to extract the time evolution of WS events is 
on-going~\cite{yabsley}; results concerning the background in the WS sample have
already been presented~\cite{belle_rws}, but none yet on $x^{\prime 2}$,
$y^\prime$. All three experiments have results for 
the time-integrated WS decay rate $R_{WS}$ \cite{CLEO_xy,BABAR_xy,belle_rws}.

\subsection{Extraction of $x^\prime$, $y^\prime$ from the time evolution of the
WS decay rate}
\noindent
Wrong-sign candidate events of the type 
$D^0 \rightarrow K^+ \pi^-$ and $\overline{D^0} \rightarrow K^- \pi^+$ 
are selected by requiring the $\pi_s$ from the $D^\star$ decay and the 
daughter $K$ of the $D^0$ to have identical 
charge (WS tag). In addition, CLEO, BABAR and Belle apply event selection 
criteria similar to those
described in Sec.~5.2, and the proper decay-time is reconstructed as
discussed in Sec.~5.1 \cite{CLEO_xy,BABAR_xy,belle_rws}. 

The measurement aims at detecting a deviation from a purely 
exponential decay law in WS events (see Sec.~3.2) by means of a 
likelihood fit to the distribution of the reconstructed proper decay-time, $t$.
The likelihood functions used by the experiments have the same principal 
structure as the ones used in 
the measurement of $y$ (see Sec.~5.3). They differentiate between a
signal and a background component and model each as the convolution of a
decay-time distribution and a resolution function. For WS events, the
decay-time distribution follows Eq.~\ref{eq:mix}, i.e. includes the three
parts discussed earlier: DCS decay, mixing and interference between the 
decays with (CF) and without (DCS) mixing.

The WS sample has only $\sim 1/300$ times the number of events of a RS sample 
selected with the same criteria as the WS sample except for the 
WS tag. This RS sample is used wherever possible to
constrain aspects in the fit that are common to the two samples.
CLEO and BABAR determine, e.g., the resolution functions of WS
and RS signals and of the common background types with the RS sample.

In the WS sample, a significant additional complication arises from the much 
lower achievable
purity\footnote{A rough estimate can be obtained by assuming that in a given 
random sample of events, the probability for a non-signal event to be 
mis-identified as a signal event is the same for WS and RS signal. Then, a 
purity of 99\% in the RS candidate sample translates into
a purity of 25\% in the WS candidate sample because WS events are 300 times 
less frequent than RS events. This simplified example does not include
background from random $\pi_s$ that together with a CF
$D^0$ decay mimic a WS event.}, 
e.g. $\sim 50\%$ in the CLEO analysis.
The fit needs to accord each background type its specific lifetime evolution,
distinct from the ones of the other background types and from those of the 
signal events. The reconstructed $m_D$ and $\delta m$ distributions provide 
a means of differentiating between the different background types and 
between background and signal.

\subsubsection{Fitting procedure}

\begin{figure}[!t] 
\centerline{\psfig{file=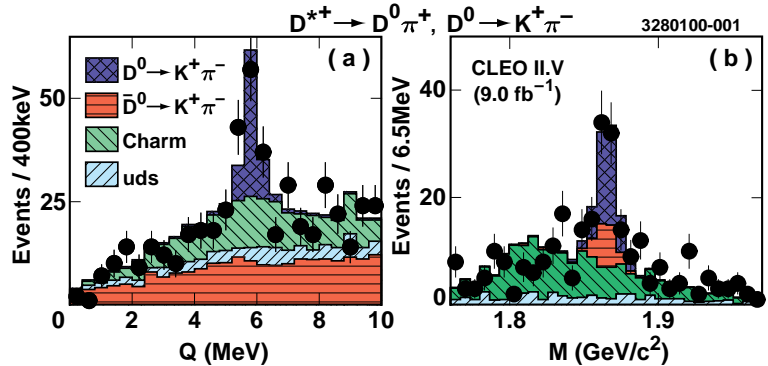,height=5cm}} 
\caption{\small Signal (dots, with statistical error bars) observed by 
CLEO~\cite{CLEO_xy} for the WS decay 
$D^0 \rightarrow K^+ \pi^-$ (charge conjugate mode included). 
The projections of the fit for the signal 
(cross-hatched) and the background (background of type A - ``uds'', 
type B - ``charm'', type C - ``$\overline{D^0} \rightarrow K^+\pi^-$'') is 
shown. $M$ (left) and $Q$ (right) are each within 2$\sigma$ of their mean 
in the RS mode. }
\label{fig:CLEO_1}
\end{figure}

\noindent
CLEO performs a two-step fitting procedure. First, the levels of the different
background types in the selected candidate sample are estimated by performing
a fit to the two-dimensional region $1.76 < m_D < 1.97$~GeV and 
$0< Q < 10$~MeV, where $Q = \delta m -m_\pi$ and $m_\pi$ is the pion 
mass.
CLEO differentiates three types of backgrounds: 
A) from the CF  
decay that together with a random $\pi_s^+$ mimics a WS event
(``$\overline{D^0} \rightarrow K^+ \pi^-$''),
B) with non-zero lifetime (``charm'') and 
C) with zero lifetime (``uds''). 
The background shapes in the $m_D - \delta m$ plane are estimated with the help of a
Monte Carlo sample that corresponds to ${\mathcal L} = 90 \ {\rm fb}^{-1}$. 
The signal shape is constrained with the help of the RS sample. 
The fit yields $\sim 45$ WS signal events in the total of
82 selected candidate events,
measured within a 2$\sigma$ window around the central values of the
$m_D$ and $Q$ distributions.   
($\sigma (m_D) = 6.4 \pm 0.1$~MeV and $\sigma (Q) = 190 \pm 2$~keV).
Figure~\ref{fig:CLEO_1} shows the observed signal as function of $Q$
and $M = m_D$ together with the projected background estimates from the fit. 
In the second step, CLEO performs a binned maximum-likelihood fit (bin size
1/20 of the $D^0$ lifetime) to the distribution of the proper 
decay-time, $t$. In this fit, the background levels from the first fit in the 
$m_D - \delta m$ plane are used.

\begin{figure}[!t] 
\begin{tabular}{cc}
\begin{minipage}[b]{.48\textwidth}
\centerline{\psfig{file=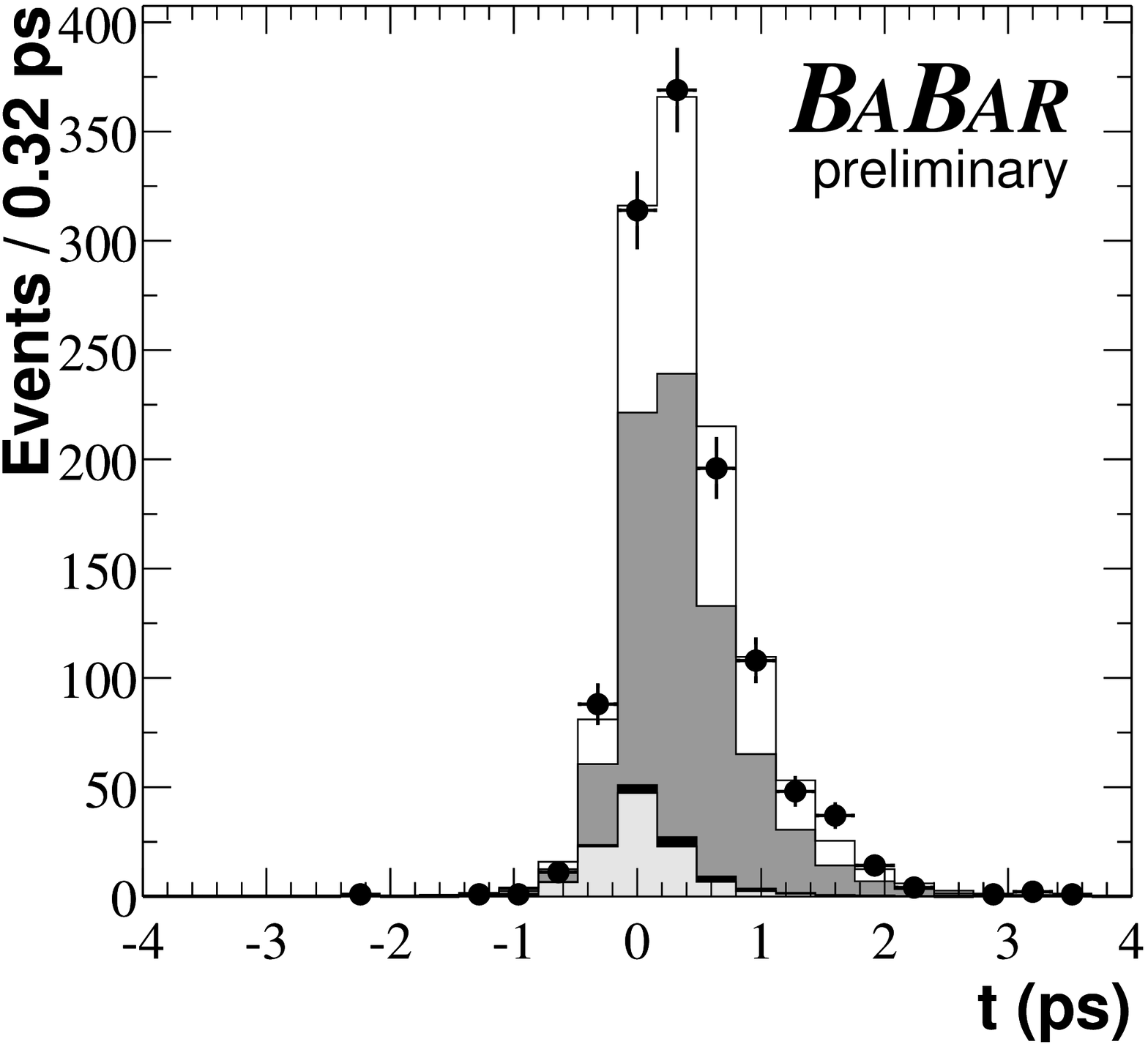,height=6cm}} 
\end{minipage}
&
\begin{minipage}[b]{.48\textwidth}
\centerline{\psfig{file=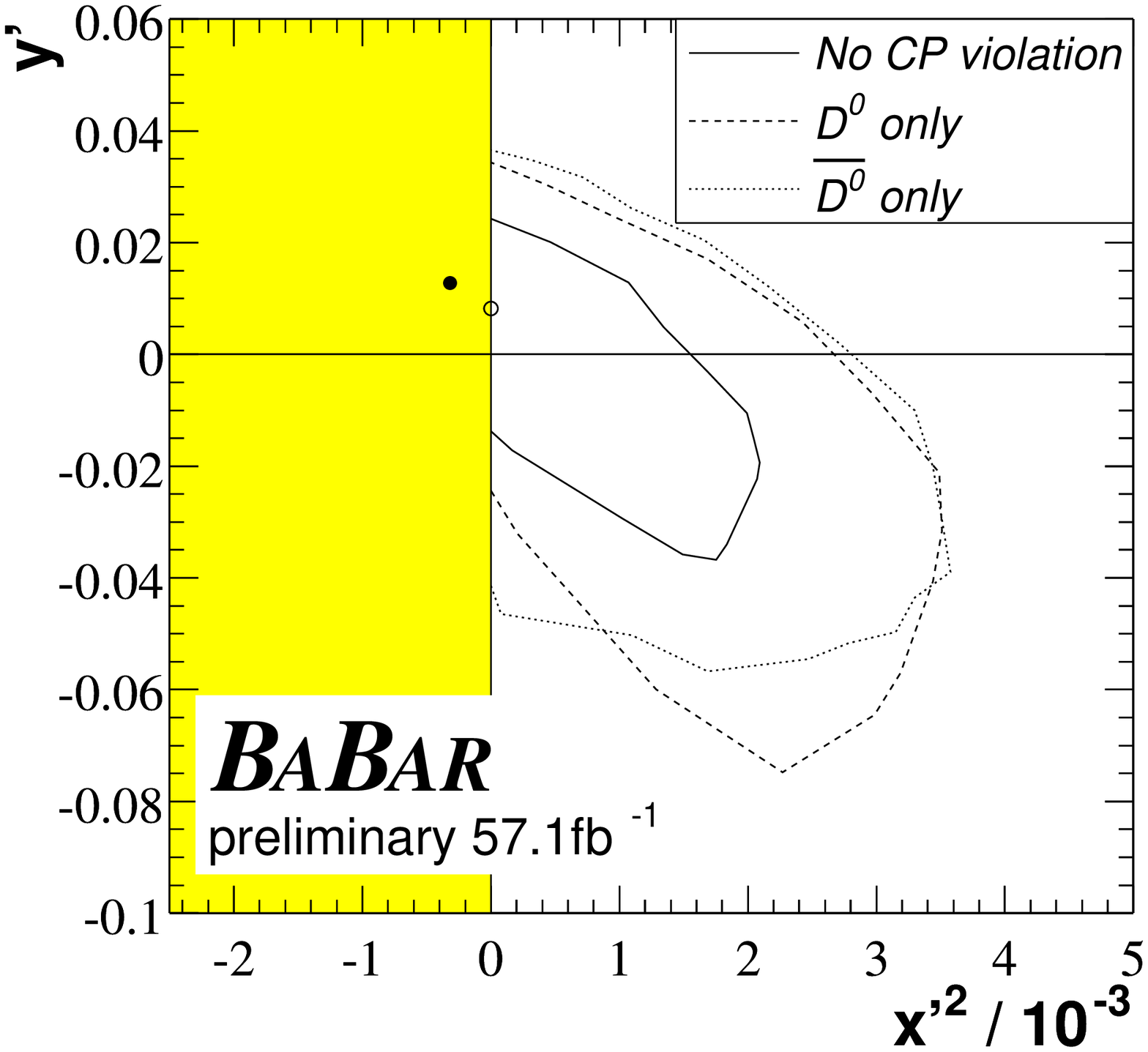,height=6cm}} 
\end{minipage}
\end{tabular}
\caption{\small (a) Distribution of the proper decay-time, $t$, in the
WS data sample (dots) compared to the fit result for the 
different background types (A: light grey, B: black, C: darker grey) and for
the signal (open). 
(b) 95\% C.L. contours, including the
systematic uncertainty, 
separately for $D^0$ 
($y^{\prime +}$ vs.~$x^{\prime + 2}$) and $\overline{D^0}$
($y^{\prime -}$ vs.~$x^{\prime - 2}$) when allowing for $CP$ violation,
and $D^0$, $\overline{D^0}$ combined ($y^\prime$ vs.~$x^{\prime 2}$)
when not allowing for $CP$ violation. The solid 
(open) point represents the most likely 
fit point in the case of no $CP$ violation without (with) the constraint 
$x^{\prime 2} > 0$.\cite{BABAR_xy}
 }
\label{fig:WSt_contours}
\end{figure}

BABAR performs an unbinned maximum-likelihood fit in several steps. First, 
the number of signal and background events is determined in a fit to the 
$m_D -\delta m$ plane. In the WS sample, BABAR models separately the 
background from A) the CF
decay that together with a random $\pi_s$ mimics a WS event, 
B) RS events where the $K$ and $\pi$ hypotheses are
swapped, C) purely combinatorial background. In the RS sample, BABAR 
considers in addition background from D) partially reconstructed $D^0$ mesons, 
typically from 3-prong $D^0$ decays.
The $\delta m$ 
distributions for background types A) and C) are obtained 
directly from the data by means of mixing a $\pi_s$ from a $D^\star$ decay of
one event with the $D^0$ candidate of another event.
The fit yields $\sim 440$ signal WS events in the region 
$1.804 < m_D < 1.924$~GeV, $\delta m < m_\pi + 25$~MeV.
Figure~\ref{fig:WSt_contours}(a) shows the reconstructed proper
decay-time distribution 
for the WS signal and the three background types A)--C).
In a second step, a fit to the RS and WS proper decay-time distributions is 
performed simultaneously to determine the resolution functions. 

Belle has reported results on the first part of the fitting procedure, the 
two-dimensional fit in the $m_D - \delta m$ plane.
Belle considers
background types similar to BABAR's.
Specifically to suppress background type B), Belle requires that 
the invariant mass of the $D^0$ candidate
calculated with swapped particle hypotheses for the daughters 
differs from the nominal mass by more than 28~MeV ($\sim 4 \sigma$). 
The remaining background types are modeled in the $m_D - \delta m$ plane 
fit function. The fit arrives at $\sim 450$ WS signal events in the region 
$1.81 < m_D < 1.91$~GeV, $0 < Q < 20$~MeV. Belle has not yet
released any results from the fit to the distribution of the proper 
decay-time, $t$.

For the WS decay-time function, CLEO and BABAR both follow
Eq.~\ref{eq:mix} and consider the case with and without $CP$ violation.
Input to all fits is $m_D$, $\delta m$, $t$, $\sigma_t$.

\subsubsection{Discussion of results}
\noindent
The results of both CLEO~\cite{CLEO_xy} and BABAR~\cite{BABAR_xy} 
are consistent with the absence of mixing and of $CP$ violation.

\begin{figure}[!t] 
\begin{tabular}{cc}
\begin{minipage}[b]{.48\textwidth}
\centerline{\psfig{file=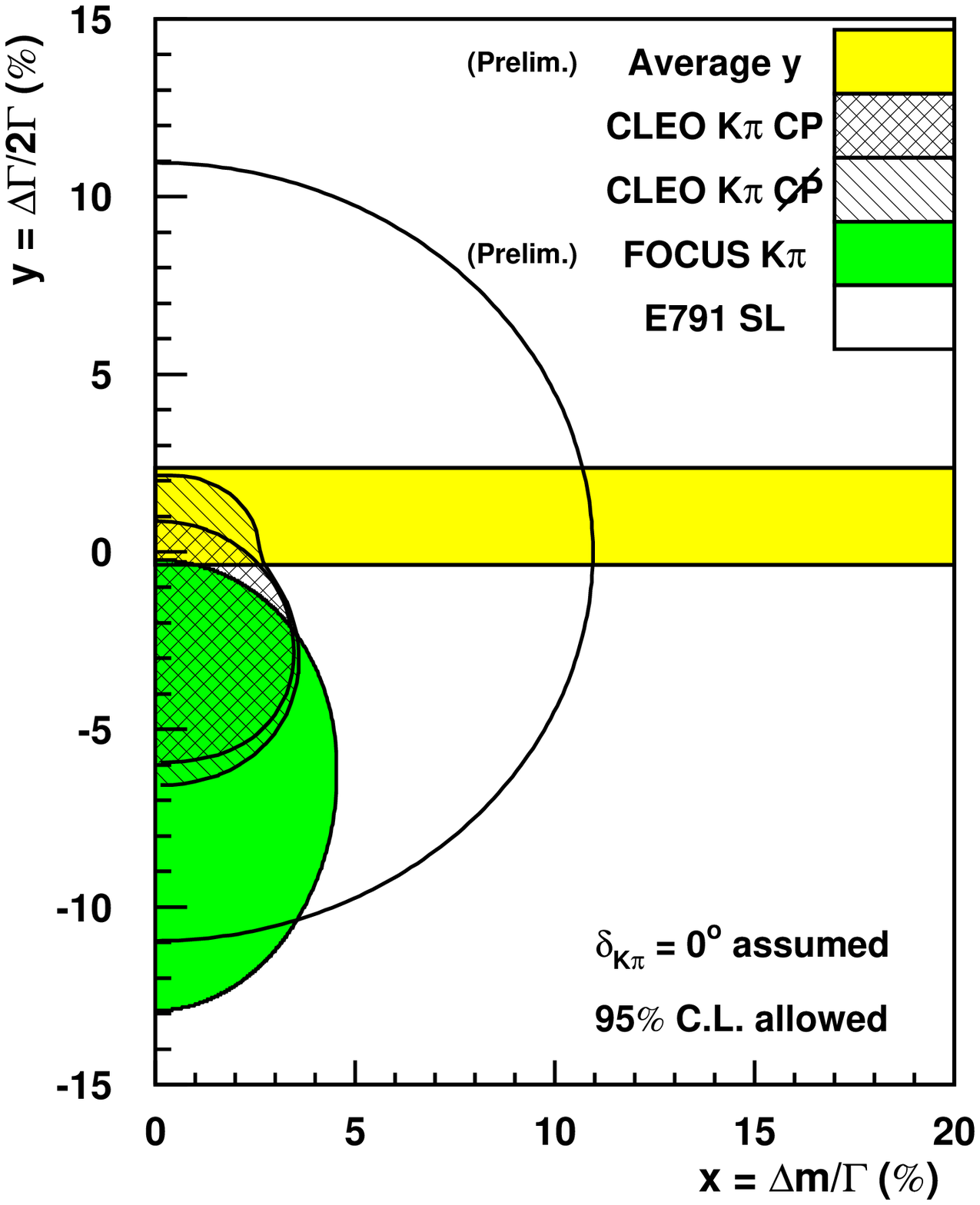,height=7.5cm}} 
\end{minipage}
&
\begin{minipage}[b]{.48\textwidth}
\centerline{\psfig{file=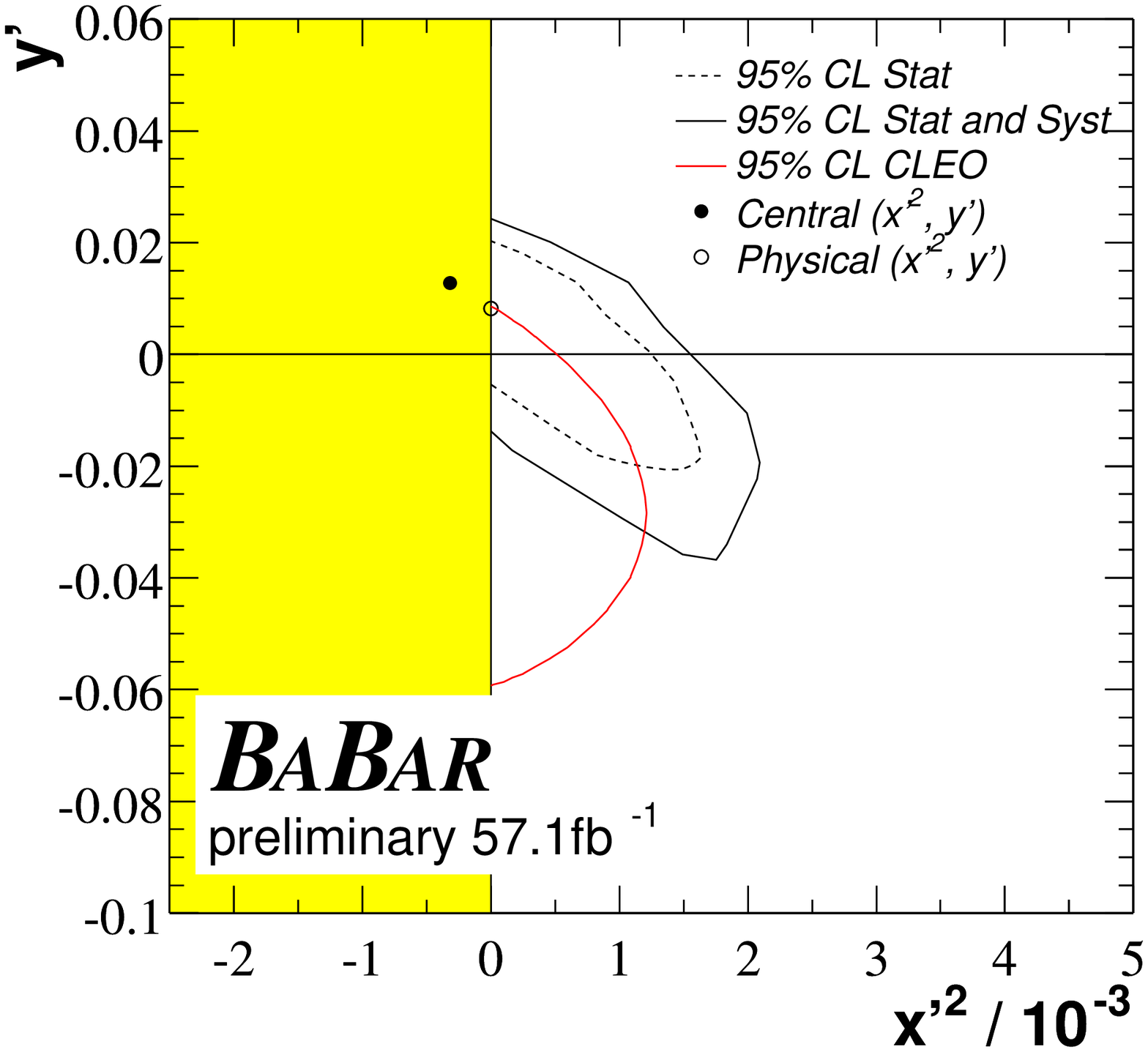,height=6cm}} 
\end{minipage}
\end{tabular}
\caption{\small (a) Comparison of the 95\% C.L. contours in the $x-y$ plane from 
E791~\cite{e791_xysl}, FOCUS~\cite{focus_xy} and CLEO~\cite{CLEO_xy}. For 
the bigger kidney-shaped (smaller) region from CLEO, $CP$ violation was (was 
not) allowed in the fit. Also shown is the band corresponding to the current 
average for $y$ obtained in lifetime-difference analyses 
(see Fig.~\ref{fig:y_comp}). The assumption $\delta_{K \pi} = 0^\circ$ for
the strong phase is made. (b) Comparison of the 95\% C.L. contours
in the $x^{\prime 2} - y^\prime$ plane from CLEO and BABAR when 
assuming $CP$ conservation in the fit. The CLEO contour does not include 
systematic uncertainties.}

\label{fig:CP_contour}
\end{figure}

CLEO constructs contours in the $x^\prime - y^\prime$ plane that contain
the true value of $x^\prime$, $y^\prime$ with 95\% confidence. Following a
Bayesian~\cite{PDG} approach, 
the 95\% confidence region includes all points in the
plane for which the negative log likelihood, $- \ln{ L}$, differs from the
best fit value by more than 3.0. All fit variables other than $x^\prime$, 
$y^\prime$ are allowed to vary to give the best fit value at each point on the
contour. The 95\% confidence region, when not and when allowing for 
$CP$ violation, is shown in Fig.~\ref{fig:CP_contour}(a). 
There, the assumption 
$\delta_{K \pi} = 0^\circ$ is made for the strong phase (see 
Eq.~\ref{eq:phase}), so that $x^\prime = x$ and $y^\prime = y$.
The contours do not include any estimate of systematic errors. 
CLEO also quotes one-dimensional limits at the 95\% C.L. for the mixing 
parameters, $(1/2) x^{\prime 2} < 0.041$\% and $ -5.8\% < y^\prime < 1.0\%$, 
without assumptions on $CP$ violation.
CLEO identifies as its dominant source of systematic error the potential 
misunderstanding of the background shapes and acceptances
and assigns a total systematic error of $\pm 0.2\%$ ($\pm 0.3\%$) for $x^\prime$
($y^\prime$).

Figure~\ref{fig:CP_contour}(a) also shows a preliminary
contour from FOCUS~\cite{focus_xy}, derived from an analysis of WS hadronic
decays of the type $D^0 \rightarrow K^+ \pi^-$, a 95\% C.L. contour 
from E791~\cite{e791_xysl}, derived from the 90\% C.L. limit 
$R_{mix} < 0.50\%$\footnote{$R_{mix} \approx 1/2(x^{\prime 2} + y^{\prime 2})
=1/2(x^2 + y^2)$.} 
in WS semileptonic decays of the type $D^0 \rightarrow K^+ l^- \nu_l$, and 
the band corresponding to the average on $y$ from the lifetime difference
analyses (see Fig.~\ref{fig:y_comp}). Not shown is the contour from 
E791~\cite{e791_rws} derived from 
the weaker 90\% C.L. limit $R_{mix} < 0.85\%$ in WS 
hadronic decays of the type $D^0 \rightarrow K^+ \pi^-$ and 
$D^0 \rightarrow K^+ \pi^- \pi^+ \pi^-$.

BABAR uses a method different from CLEO's to arrive at a 95\% C.L. contour.
BABAR argues that due to allowing $x^{\prime 2}$ in its fit to take unphysical
negative values, it is not clear how to apply a Bayesian ansatz to derive
an error estimate from the two-dimensional likelihood distribution. In 
addition, BABAR finds that the likelihood distribution depends strongly
on the most likely fitted values of $x^{\prime 2}$ and $y^\prime$. 
To define 95\% C.L. contours, BABAR applies a frequentist approach based on 
toy Monte Carlo (MC) experiments. Any point 
$\vec{\alpha_c} = (x_c^{\prime 2}, y^\prime_c)$ on the 95\% C.L. contour has
to meet the requirement: If a
toy MC experiment is generated at that point, there is a 95\% probability that 
the ratio 
$\Delta \ln{L} (\vec{\alpha_c}) = \ln{L} (\vec{\alpha_c})
- \ln{L_{max}}$ is greater than 
$\Delta \ln{L_{data}}(\vec{\alpha_c})$ calculated for the data.
There, $L_{max}$ is the maximum likelihood obtained in the fit 
either to data or to a toy MC sample. 
BABAR constructs 95\% C.L. contours also for the systematic effects considered
in the analysis, among them uncertainties in the form of the fit functions,
detector effects and effects of the selection criteria. 
Figure~\ref{fig:WSt_contours}(b) shows the resulting 95\% C.L. contours that 
include the statistical as well as the systematic uncertainty estimate.
A strong correlation between $x^{\prime 2}$ and $y^\prime$ is apparent.
The most likely fit point in the case of no $CP$ violation has a negative
coordinate in $x^{\prime 2}$. 

A direct comparison of the CLEO and BABAR results is not possible for the
case when $CP$ violation is allowed in the fit. CLEO uses as fit output
parameters $x^\prime$, $y^\prime$, $R_D$ and $A_D$, $A_M$, $\sin{\phi}$,
while BABAR uses $x^{\prime + 2}$, $y^{\prime +}$, $R^+$ 
($x^{\prime - 2}$, $y^{\prime -}$, $R^-$) for the $D^0$ ($\overline{D^0}$) 
case (see Sec. 3.2). 

\begin{table}[!t]
\caption{\small Comparison of the 95\% C.L. limits for the fit output parameters
of BABAR and CLEO when $CP$ conservation is assumed in the fit. BABAR's limits
include systematic uncertainties and were obtained in a fit that allowed 
$x^{\prime 2 } < 0$.}
\centerline{\footnotesize }
\centerline{
\begin{tabular}{|c||c| c| c| c| }
\hline
& $R_D$ & $y^\prime$ & $x^\prime$ & $x^{\prime 2}$ \\ \hline
CLEO [\%]&(0.24, 0.69) & (-5.2, 0.2) & (-2.8, 2.8) & $ <0.076$  \\
BABAR [\%]&(0.22, 0.46) & (-3.7, 2.4) & -  & $<0.21$  \\
\hline
\end{tabular}}
\label{tab:xy_comp}
\end{table}

The results when assuming $CP$ conservation in the fit are in principle 
comparable between CLEO and BABAR, see Tab.~\ref{tab:xy_comp}. BABAR includes
systematic uncertainties in its 95\% C.L. contour and obtains its limits
on $x^{\prime 2}$ and $y^\prime$ by projecting this contour onto the 
corresponding axis. 
The CLEO limits in Tab.~\ref{tab:xy_comp}, instead,
correspond to one-dimensional 95\% C.L. intervals, determined by an increase in $- \ln{L}$ of 1.92 compared to the best
fit value, and do not include systematic uncertainties.

BABAR's upper limit on $x^{\prime 2}$ is almost three times 
bigger than CLEO's, in spite of being based on a six times larger data 
sample. A possibly overly conservative estimate of the systematic error
cannot account for a difference of this size, as illustrated by 
Fig.~\ref{fig:CP_contour}(b). There, the CLEO 95\% C.L. contour
(no systematic errors) is overlaid with the two BABAR contours that
are obtained before and after adding the systematic uncertainty. 
Two possible reasons for so pronounced a difference between
CLEO and BABAR are the different techniques for obtaining the 95\% C.L. 
limits and the treatment of the fit output
parameter $x^\prime$. BABAR allows $x^{\prime 2}$ to take unphysical negative
values in its fit, while CLEO excludes this possibility by choosing $x^\prime$
instead of $x^{\prime 2}$ as fit parameter.
Another possible reason is the sign of the fit result on $y^\prime$. Toy 
Monte Carlo studies indicate that the 95\% C.L. contour differs in size and 
shape depending on the sign \cite{yabsley,Ulrik_priv}. For positive 
$y^\prime$ (the BABAR case), the contour tends to be larger than for negative
$y^\prime$ (the CLEO case). Albeit its precise origin is not yet understood,
this behavior may point to a qualitative difference between the two regimes, 
with destructive (constructive) interference between the decays with (CF) and 
without (DCS) mixing for $y<0$ ($y>0$).

\subsection{Extraction of the time-integrated WS decay rate $R_{WS}$} 
\noindent
CLEO and BABAR arrive at a measurement of $R_{WS}$ by repeating the fits 
described above with the assumption of no mixing in the $D^0$ system, i.e. 
$x = y = 0$ \cite{CLEO_xy,BABAR_xy}.

Belle uses its fit in the $m_D - \delta m$ plane, described above, to determine
the time-integrated number of signal events in the candidate 
samples \cite{belle_rws}.
The ratio of the number of signal events in the WS and RS candidate samples 
yields $R_{WS}$. 
The systematic error on $R_{WS}$ is dominated by the uncertainty on the
background shapes used in the fit.

\begin{figure}[!t] 
\centerline{\psfig{file=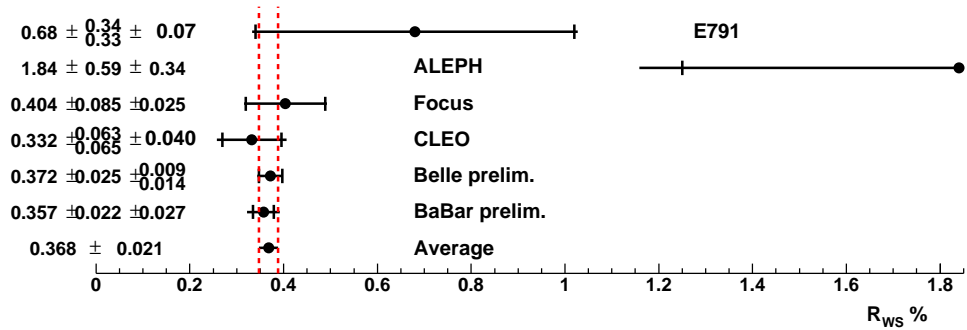,height=4.8cm}} 
\caption{\small Comparison of results for $R_{WS}$ from different 
experiments \cite{CLEO_xy,BABAR_xy,belle_rws,e791_rws,aleph_rws,focus_rws}.
The total size
of the error bars corresponds to the quadratic sum of statistical and 
systematic error, the inner part indicates the size of the statistical 
error only. The average is calculated as mean of the individual
measurements, each weighted by the quadratic sum of its statistical and
systematic error.}
\label{fig:RWS_comp}
\end{figure}

Figure~\ref{fig:RWS_comp}  compares the results from CLEO~\cite{CLEO_xy}, based
on ${\mathcal L} = 9.0 \ {\rm fb}^{-1}$, BABAR~\cite{BABAR_xy} 
($57.1 \ {\rm fb}^{-1}$) and Belle~\cite{belle_rws} ($46.2 \ {\rm fb}^{-1}$)
to earlier 
measurements by E791~\cite{e791_rws}, ALEPH~\cite{aleph_rws} and 
FOCUS~\cite{focus_rws} in WS decays of the type $D^0 \rightarrow K^+ \pi^-$. 

In the absence of mixing, $R_{WS}$ corresponds to the the doubly
Cabibbo-suppressed decay rate $R_D$ (see Sec.~3.2). 
The discrepancy between the SM 
expectation in the absence of mixing, $R_D \approx 0.25$\%,
and the values of $R_{WS}$ measured by CLEO, BABAR and Belle may be 
attributable to the effect of SU(3) symmetry breaking \cite{nir}. 

\section{Summary and Outlook}
\noindent
The asymmetric B-factories PEP-II and KEK-B with the experiments
BABAR and Belle, operational since 1999, have rendered possible studies of 
mixing in the $D^0$ system with unprecedented statistical precision and 
sample purities.
First results for the mixing parameters 
$x$ and $y$ are compatible with an absence of mixing and of $CP$ 
violation. 

BABAR and Belle both expect to reach
${\mathcal L} = 500~{\rm fb}^{-1}$ by 2006. In data sets of this size,
the statistical errors in the measurements reviewed in this article 
can be reduced by a factor
of three. For $y$, a statistical precision of $\sim 0.2\%$ 
would be within reach and could be further improved by, e.g., employing
additional $D^0$ decay channels of well-defined $CP$ symmetry and, for BABAR,
by investigating the possibility of dropping the $D^\star$ tag requirement
in its event selection.

BABAR's preliminary result for 
$x^{\prime 2}$, $y^\prime$,
determined from the time evolution of WS decays of the type 
$D^0 \rightarrow K^+ \pi^-$, is the second such
result from a collider experiment after CLEO's. Already with its present
data sample, BABAR should be able to improve its systematic error 
estimate substantially. Future results from Belle should help shedding light
on the differences between the CLEO and the BABAR result. 
Additional information may be gained from other WS $D^0$ decay 
channels, hadronic and semileptonic ones, which have already been observed 
in other experiments (see e.g. Ref.~\cite{e791_rws,cleo_otherWS}).

Given the huge uncertainties in Standard Model (SM) predictions for 
$D^0$ mixing, it might prove difficult to establish physics beyond the SM
from the size of the measured $x$ and $y$ parameters alone. A more
robust potential signal for new physics may well be $CP$ violation in the
$D^0$ system \cite{petrov}. Efforts to establish $CP$ violation in the
$D^0$ system are already part of the WS time evolution analyses discussed
in this review and are likely to intensify in the future.

Methods for investigating mixing in the $D^0$ system
complementary to those at the B-factories would be available at CLEO-c.
CLEO-c at CESR-c, currently under discussion as successor to CESR at Cornell,
may operate at the $D^0 \overline{D^0}$ threshold. There, the
quantum-mechanical coherence of the produced  $D^0$, $\overline{D^0}$ pair
can be exploited to study mixing in the $D^0$ system in ways not
available\footnote{The coherent state of the $D^0$, $\overline{D^0}$ pair
renders possible, for example, a direct measurement of the strong phase
$\delta_{K \pi}$ (see Eq.~\ref{eq:phase}) between the DCS and CF 
decays \cite{cleo-c,cleo-c2}.}  
to any already existing experiment \cite{cleo-c,cleo-c2}.

\vspace{15pt}
\noindent
{\bf \Large Acknowledgments}
\vspace{12pt}

\noindent 
I would like to thank Ulrik Egede and Bruce Yabsley for enlightening 
discussions, helpful suggestions and feedback on this review. The help of the BABAR,
Belle, CLEO and FOCUS collaborations by making available several figures
for inclusion in this review is gratefully acknowledged, as is the help of
Harry Cheung in preparing Fig.~6(a). I would also like to thank
Tom Browder, Pat Burchat, Nick Ellis, Brian Meadows, Yossi Nir, Abe Seiden and 
David Williams for their comments on the manuscript.

\end{document}